\documentclass[%
 reprint,
nofootinbib,
 amsmath,amssymb,
 aps,
pra,
]{revtex4-2}
\usepackage{lineno,hyperref}
\usepackage[caption=false]{subfig}
\usepackage[nolist]{acronym}
\usepackage{graphicx}
\usepackage{dcolumn}
\usepackage{bm}
\usepackage{xcolor}
\usepackage[normalem]{ulem}

\acrodefplural{AOM}[AOMs]{acousto-optic modulators}
\acrodefplural{FSR}[FSRs]{free spectral ranges}
\DeclareUnicodeCharacter{2212}{-}
\begin{acronym}[NRLO] 

\acro{AOM}{acousto-optic modulator}
\acro{ASD}{amplitude spectral density}
\acro{BFRT}{Brookhaven\,$|$\,Fermilab\,$|$\,Rochester\,$|$\,Trieste}
\acro{BMV}{Birefringence Magnétique du Vide}
\acro{DESY}{Deutsches Elektronen-Synchrotron}
\acro{EOM}{electro-optic modulator}
\acro{FWHM}{full width half maximum}
\acro{FSR}{free spectral range}
\acro{HERA}{Hadron–Electron Ring Accelerator}
\acro{HWP}{half-wave plate}
\acro{OVAL}{Observing VAcuum with Laser}
\acro{PD}{photodetector}
\acro{PDH}{Pound-Drever-Hall}
\acro{PVLAS}{Polarization of the Vacuum with LASer}
\acro{QED}{quantum electrodynamics}
\acro{RAM}{residual amplitude modulation}
\acro{VMB}{Vacuum magnetic birefringence}
\end{acronym}

\begin{document}

\preprint{APS/123-QED}

\title{Demonstration of an interferometric technique for measuring vacuum magnetic birefringence with an optical cavity }

\author{Aaron~D.~Spector}
 \email{aaron.spector@desy.de}
\author{Todd~Kozlowski}%
 \affiliation{%
Deutsches Elektronen-Synchrotron DESY, 22603 Hamburg, Germany}%
\author{Laura~Roberts}
\affiliation{
Max-Planck-Institut f\"ur Gravitationsphysik (Albert-Einstein-Institut) \protect\\ and Leibniz Universit\"at Hannover, 30167 Hannover, Germany}%


\date{\today}

\begin{abstract}
    Vacuum magnetic birefringence (VMB) is an effect predicted by quantum electrodynamics, in which the vacuum behaves as a non-linear optical medium and exhibits a birefringence in the presence of a magnetic field. In this work, an interferometric scheme is introduced to measure this effect for the first time by sensing the changes in the frequencies of fields stabilized to the resonances of an optical cavity. Results are presented from a prototype setup with this novel sensing technique implemented on a 19\,m test cavity without a magnetic field. We propose using this scheme for a measurement of VMB with a 245\,m long optical cavity and a string of 24 superconducting magnets, arranged for the ALPS\,II experiment. In this manuscript  we examine potential sources of noise in the prototype, and project these results in terms of the sensitivity of the full-scale experiment.
\end{abstract}

\maketitle

\acrodef{AOM}{acousto-optic modulator}
\acrodef{ASD}{amplitude spectral density}
\acrodef{BFRT}{Brookhaven\,$|$\,Fermilab\,$|$\,Rochester\,$|$\,Trieste}
\acrodef{BMV}{Birefringence Magnétique du Vide}
\acrodef{DESY}{Deutsches Elektronen-Synchrotron}
\acrodef{EOM}{electro-optic modulator}
\acrodef{FWHM}{full width half maximum}
\acrodef{FSR}{free spectral range}
\acrodef{HERA}{Hadron–Electron Ring Accelerator}
\acrodef{HWP}{half-wave plate}
\acrodef{OVAL}{Observing VAcuum with Laser}
\acrodef{PD}{photodetector}
\acrodef{PDH}{Pound-Drever-Hall}
\acrodef{PVLAS}{Polarization of the Vacuum with LASer}
\acrodef{QED}{quantum electrodynamics}
\acrodef{RAM}{residual amplitude modulation}
\acrodef{VMB}{Vacuum magnetic birefringence}

\section{Introduction}

\ac{VMB} describes how, in the presence of a magnetic field, vacuum will exhibit an increase in its index of refraction that depends on the polarization of light relative to the direction of the magnetic field. Following Dirac's theory of the positron \cite{Dirac}, \ac{VMB} was predicted by the Euler-Heisenberg Lagrangian via virtual electron-positron pairs  \cite{euler1935streuung,heisenberg1936folgerungen}, outlined in the theory of \ac{QED}. Although a host of experiments have attempted to observe \ac{VMB} over the last 40 years, a direct measurement confirming its existence has proven elusive. 
This is mainly due to the interaction strength being so small, with standard \ac{QED} predicting changes in the index of refraction for polarization axes parallel ($\Delta n_{\parallel}$) and perpendicular ($\Delta n_{\perp}$) to the magnetic field direction of 
\begin{align}
    &\Delta n_{\parallel}=7A_eB^2 {\hspace{1cm}\rm and}\\
    &\Delta n_{\perp}=4A_eB^2
\end{align}
where $A_e = 1.32 \times 10^{-24}\,\rm T^{-2}$ is a constant representing the strength of these nonlinear effects \cite{heisenberg1936folgerungen}. This leads to a differential index of refraction between the polarization axes, also known as birefringence, as a function of the magnetic field strength,
\begin{equation}
  \Delta n_{_{\rm VMB}}  = \Delta n_{\parallel} - \Delta n_{\perp}=3A_eB^2.
  \label{Eq:Del_n_vmb}
\end{equation}
For a 5.3\,T magnetic field, $\Delta n_{_{\rm VMB}} \simeq 10^{-22}$.

The prospect of measuring a deviation from the \ac{QED} predicted amplitude of $\Delta n_{_{\rm VMB}}$ or an absence of the effect is perhaps even more exciting, as this would serve as evidence for new physics. Hypothetical particles beyond the standard model, for example 
millicharged particles, can also induce polarization effects \cite{Gies:2006ca}, such that sensitive measurements of the quantum vacuum serve as probes of these theories \cite{Marmier:2024rwd}. 

A typical strategy to measure \ac{VMB} is to pass a laser through a magnetic field and observe its change in polarization state. The first measurements of this kind were performed by the \ac{BFRT} collaboration using an optical delay line to increase the interaction length between the laser and the magnetic field \cite{cameron1993search}. Modern experiments now use optical cavities to further extend the optical path length of the laser through the magnetic field. 
It is also essential to modulate the magnetic field to distinguish the signal from noise, with higher modulation frequencies typically corresponding to better sensitivities \cite{ejlli2020pvlas}. The modulation can be done in a number of ways. \ac{BFRT} performed the modulation of the magnetic field by ramping the current of a superconducting dipole magnet, achieving a modulation frequency of 32\,mHz. The \ac{PVLAS} experiment opted to rotate permanent magnets \cite{ejlli2020pvlas} to increase the signal frequency to 16\,Hz, while other experiments such as \ac{BMV} and \ac{OVAL} utilized pulsed magnets to achieve even higher characteristic frequencies above 100\,Hz \cite{hartman2019characterization,fan2017oval}. The most sensitive measurement to date was from \ac{PVLAS} with a sensitivity of $\Delta n_{_{\rm VMB}}  < 3\times10^{-23}\rm\,T^{-2}$, less than a factor of 8 from the \ac{QED} prediction.

An alternative to the polarimetric approach was proposed in Ref.~\cite{hall2000measurement}. Three laser fields are stabilized to three resonances of an optical cavity with the center resonance polarized orthogonally to the magnetic field and the outer two resonances polarized parallel to it. 
The \ac{VMB} effect can then be measured by observing the relative frequency changes between the two outer resonances and the center resonance when the magnetic field is modulated.  In this case, the finesse of the optical cavity does not directly amplify the signal, but is used to improve the precision with which the lasers can be stabilized to the cavity resonances. By using three resonances, it is also possible to decouple the measurement from the length noise of the cavity. 


This paper provides a demonstration of the sensing scheme for this technique using a 19\,m optical cavity without a magnetic field. A heterodyne detection system is used to read out the frequency changes between the cavity resonances. 
The ultimate goal of this prototype setup is to demonstrate that this technique is capable of achieving a sensitivity sufficient to measure \ac{VMB} using the magnet string assembled for the Any Light Particle Search II (ALPS\,II) at the \ac{DESY}, in Hamburg, Germany. 

\begin{figure}[t]
    \centering 
   \includegraphics[width=0.9\linewidth]{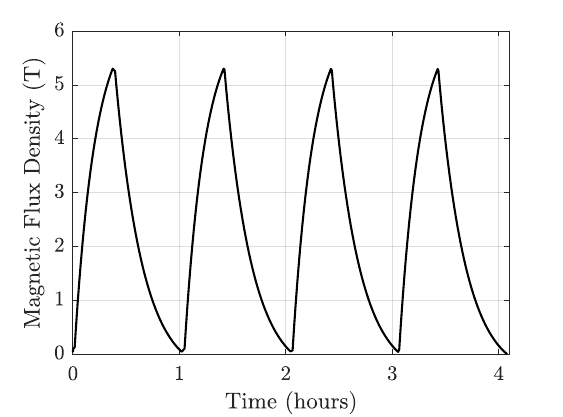}
    \caption{Modulated magnetic flux density of the ALPS\,II magnetic string. 
    \label{fig:mag_T}}
\end{figure}

In ALPS\,II, 24 superconducting dipole magnets formerly used by the \ac{HERA}, each generating a 5.3\,T dipole field over 8.8\,m of length, have been straightened and aligned to give a total $B^2 L$ of 5990\,$\rm T^2\, m$ \cite{albrecht2021straightening}. This corresponds to a predicted optical path difference of $2.37\times10^{-20}$\,m between fields polarized parallel and perpendicular to the direction of the magnetic field. Comparing this to the magnets used in \ac{PVLAS} with a $\Delta B^2 L$ of 10\,$\rm T^2\, m$, the ALPS\,II  magnet string would therefore be able to produce a \ac{VMB} signal nearly 600 times larger than the most sensitive \ac{VMB} experiment to date. A characteristic modulation frequency of 0.3\,mHz was demonstrated in the ALPS\,II magnet string by ramping and discharging the current to generate 4 cycles over the course of 4 hours, as shown in Figure~\ref{fig:mag_T}. It may be possible to increase this frequency up to 5\,mHz with upgrades to the system. Such low modulation frequencies present a challenge, as environmental noise tends to rapidly increase at lower frequencies \cite{ejlli2020pvlas}.

ALPS\,II has shown it is possible to operate a high-finesse optical cavity whose eigenmode propagates through the bore of the magnet string \cite{kozlowski2024designperformancealpsii}. The intrinsic birefringence noise of the cavity itself may also limit the sensitivity of the experiment at these frequencies, as has likely been the case in several previous \ac{VMB} experiments \cite{ejlli2020pvlas}.
Current literature on the intrinsic birefringence noise of optical cavities appears to show that it has an \ac{ASD} that is inversely proportional to both the square root of the Fourier frequency and the $1/e^2$ beam radius ($w$) of the cavity eigenmode on the mirror surface  \cite{ejlli2020pvlas,YANG2025111660,PhysRevX.13.041002}. The noise levels measured by previous experiments indicate that the intrinsic birefringence noise of the cavity will lead to an incoherent noise floor in our sensitivity to differential optical path length changes at a level of
\begin{equation}
    \tilde{\delta L}_\theta \sim \left(\frac{\rm 9\,mm}{w}\right) \cdot\sqrt{\left(\frac{\rm 3\,mHz}{f}\right)} \cdot10^{-17}\rm\,m/\sqrt{Hz}.
\label{Eq:Bi_noise}
\end{equation}
Here, the relative uncertainty in the scaling of this noise floor is roughly 50\% due to the range of values across the measurements in these papers. For a signal frequency at 3\,mHz and a beam radius on the cavity mirrors of 9\,mm, this equation predicts a differential path length noise of $10^{-17}\rm\,m/\sqrt{Hz}$. Nevertheless, if the measurement is limited by cavity birefringence noise,  a signal-to-noise ratio of unity could be achieved for a signal strength at the \ac{QED}-predicted level after less than two days of integration time by cross-correlating the expected birefringence signal, due to the magnet modulation, with the measured data. Therefore, the outlook for a measurement of \ac{VMB} with the ALPS\,II magnet string is promising.

This paper is organized as follows. In Section~\ref{Sec:3_res} the design of the optical system is discussed and in Section~\ref{Sec:Proto} the experimental setup, implemented on a 19\,m long cavity, is presented. The results of measurements using this system are given in Section~\ref{Sec:Results}. This includes a measurement of the noise of the setup as well as the static birefringence of the optical cavity. In Section~\ref{Sec:Conc} conclusions from the experiment are discussed along with the prospects of measuring \ac{VMB} with a full-scale experiment using a 245\,m cavity and the ALPS\,II magnet string.

\section{Optical System}
\label{Sec:3_res}

The optical system is designed to measure changes in the birefringence of a cavity via the differential frequency changes between laser fields. To achieve this, three separate linearly polarized fields are stabilized to different resonances of the cavity using the \ac{PDH} laser frequency stabilization technique \cite{pound1946electronic,drever1983laser,black2001introduction}. A laser on resonance with an optical cavity propagates for some integer number $N$ of wavelengths over a round trip through the cavity. This leads to the resonance condition,
\begin{equation}
    2nkL+\theta_{_{\rm M}} = 2\pi N.
\end{equation}
In this equation $n$ is the index of refraction of the medium between the mirrors, $k$ is the wave vector of the laser, $L$ is the single pass distance between the mirrors, and $\theta_{_{\rm M}} $ is the additional round trip phase accumulated in the mirror coatings. 

The frequency difference between two lasers with orthogonal polarization states that are stabilized to  the cavity $\Delta N$ resonances apart can be expressed as (assuming $\Delta N/ N \ll 1$)
\begin{equation}
    \Delta\nu\simeq f_0(N\Delta n + \Delta N - \Delta\theta/2\pi )
    \label{Eq:Del_nu}.
\end{equation}
In the above equation $\Delta n$ is the difference in the indices of refraction experienced by the orthogonally polarized fields as they propagate between the cavity mirrors and we assume $\Delta n \ll n$. For simplicity, we also assume that the birefringent medium spans the entire distance between the mirrors. The \ac{FSR} of the cavity is given by $f_0=c/2nL$ and therefore the laser frequency $\nu$ is equal to $Nf_0$. 
The term $\Delta\theta=\theta_{\parallel}-\theta_{\perp}$ gives the difference in the phase accumulated in the mirror coatings for the orthogonal polarization states. This is due to the birefringence of the dielectric mirror coatings and will lead to an offset in the difference frequency between the lasers from some integer multiple of \ac{FSR}s by $\Delta \nu_\theta=f_0\Delta\theta/2\pi$. Fluctuations in the differential reflected phase from the mirror coatings will lead to birefringence noise in the cavity. This equation also assumes no dispersion effects in the coatings. A calculation of the effects related to dispersion can be found in Appendix~\ref{App:Dis}.


If there is some birefringent medium between the mirrors 
this will appear as an offset in the difference frequency between orthogonally polarized lasers of $\Delta\nu_n = \nu\Delta n$. 
The birefringence signal generated by \ac{VMB} at the magnet modulation frequency, $\Delta n_{_{\rm VMB}}$ in Equation~\ref{Eq:Del_n_vmb}, will introduce a modulation in the frequency difference between the resonances of the cavity for the orthogonally polarized fields of 
\begin{equation}
\Delta \nu_{_{\rm VMB}} = \nu \,\Delta n_{_{\rm VMB}}\frac{L_{_B}}{L}.     
\end{equation}
Here the length of the magnetic field $L_{_B}$ must be distinguished from the length of the cavity $L$. This can also be expressed as a differential change in the cavity length of $\Delta L_{_{\rm VMB}}= L_{_B}\Delta n_{_{\rm VMB}}$ experienced by the orthogonal polarization states. It should be noted that in this manuscript lengths such as $L$,  $\Delta L_{_{\rm VMB}}$, and $\delta L$ will refer to the single pass lengths (the physical distance between the mirrors) and their changes, respectively.


Changes in the length of the cavity will cause 
fluctuations in its \ac{FSR}, $f_0$ in Equation~\ref{Eq:Del_nu}. This will lead to additional noise in the frequency difference between the lasers of $\delta \nu_L=\Delta N \, f_0 \, \delta L/L$. Therefore, if the frequency difference between the lasers ($\Delta N \, f_0$) is 5\,MHz for a \ac{VMB} measurement using the  ALPS\,II magnet string with a 245\,m cavity, length noise of 1\,{\textmu}m/$\sqrt{\rm Hz}$ (typical length noise of the 123\,m cavity used in ALPS\,II \cite{kozlowski2024designperformancealpsii}) would produce noise in the differential cavity length measurement of $1.8\times 10^{-14}$\,m/$\sqrt{\rm Hz}$, making an observation of the \ac{VMB} effect practically impossible.

\begin{figure}[t]
    \centering 
  \subfloat[ Frequency configuration of the three laser fields \label{fig:freq}]{
  \includegraphics[width=0.9\linewidth]{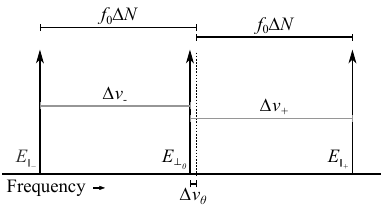}
    }
     
    \subfloat[Cavity birefringence versus VMB \label{fig:freq_zoom}]{
    \includegraphics[width=0.9\linewidth]{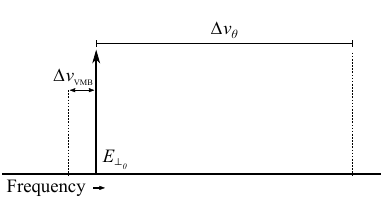}
    }
    \caption{In (a) the frequencies of the fields in the birefringence measurement scheme at the detection port of the cavity are shown. The fields $E_{\parallel_-}$, $E_{\perp_0}$, and $E_{\parallel_+}$ are all stabilized to different cavity resonance separated by a fixed number of \ac{FSR}s. In (b) we enlarge the frequency axis to show that the \ac{VMB} signal, $\Delta\nu_{_{\rm VMB}}$, manifests itself as a frequency change of the beatnote in the same way as the cavity birefringence $\Delta\nu_{_{\theta}}$.}
    \label{fig:freq_total}
\end{figure}

An additional field sensing the frequency changes of a third resonance can be used to cancel the coupling of the cavity length noise to the measurement. First, a field $E_{\perp_0}$, polarized perpendicular to the direction of the magnetic field, is stabilized to a resonance of the cavity. Two other fields $E_{\parallel_-}$ and $E_{\parallel_+}$, with polarization states aligned to the magnetic field, are then stabilized to cavity resonances some integer number of \ac{FSR}s, $\Delta N$, above and below the resonance occupied by $E_{\perp_0}$. Figure~\ref{fig:freq} shows a diagram of the configuration of laser frequencies in the setup. 
Here, looking at the frequency of the interference beatnote between $E_{\parallel_-}$ and $E_{\perp_0}$, given by $\Delta\nu_-$, and that of $E_{\parallel_+}$ and $E_{\perp_0}$, given by $\Delta\nu_+$, it is apparent that changes in the \ac{FSR} will be common to both $\Delta\nu_+$ and $\Delta\nu_-$. The coupling of changes in $f_0$ to the birefringence measurement can therefore be drastically reduced by subtracting $\Delta\nu_-$ from $\Delta\nu_+$, as in
\begin{equation}
    \frac{\Delta\nu_+ - \Delta\nu_-}{2} \simeq \Delta\nu_{_{\rm VMB}}  - \Delta\nu_\theta + \delta\nu_{_{\theta}} .
\label{Eq:sup_beat}
\end{equation}
In the equation above, the term $f_0\Delta N$ from Equation~\ref{Eq:Del_nu} has been canceled. The term $\Delta\nu_{_{\rm VMB}}$ is the birefringence signal induced by \ac{VMB}, while $\Delta\nu_\theta$ and $\delta\nu_{_{\theta}}$ are, respectively, the static and dynamic birefringence of the cavity.

The differential optical path length changes for the orthogonally polarized states be expressed as
\begin{equation}
    \Delta L_n\simeq \frac{L}{2}\frac{\Delta\nu_+ - \Delta\nu_-}{\nu} 
    \label{Eq:f_to_dL}.
\end{equation}
Then,
\begin{equation}
    \Delta n_{_{\rm VMB}} \simeq \frac{\Delta L_n}{L_{_B}}
    \label{Eq:del_f}
\end{equation}
can be used to express the birefringence induced by \ac{VMB} in terms of the optical path length changes at the magnet modulation frequency. In this work, the laser frequency is $\nu=282\rm\,THz$. If sufficient precision in the laser frequency stabilization loops can be achieved, the sensitivity of the system will likely be limited by the intrinsic birefringence noise of the cavity,  $\delta L_\theta$ from Equation~\ref{Eq:Bi_noise}. Nevertheless, projections based on the results from measurements of other high-finesse optical cavities \cite{ejlli2020pvlas,YANG2025111660,PhysRevX.13.041002} suggest that this noise should be at most $10^{-17}\rm\,m/\sqrt{Hz}$ at 3\,mHz for a 245\,m cavity with a 9\,mm beam radius on the mirrors. This would allow the system to measure the differential optical path length changes induced by the \ac{VMB} effect at the amplitude predicted by \ac{QED}, with a signal-to-noise ratio of 3 after roughly 1,600,000 seconds of integration time.

\section{Prototype Setup}
\label{Sec:Proto}

One of the primary challenges in using this technique is the very precise stabilization of the three laser fields to their cavity resonances. If a precision in the control system of $10^{-18}\rm \,m/\sqrt{Hz}$, in terms of $\Delta L_n$,  is targeted for a 245\,m cavity with a finesse of 100,000,\footnote{The finesse of the 245\,m cavity is expected to be limited by scattering losses due to the surface roughness of the substrate over the beam spot area, which will have a $1/e^2$ radius of 9\,mm. A finesse of 100,000 lies at the limit of what is possible with state of the art polishing techniques for beams of this size \cite{Isogai:13,Drori:22,kozlowski2024designperformancealpsii}.} this would require stabilizing the laser frequencies to $0.2\rm\,ppm/\sqrt{Hz}$ of the cavity \ac{FWHM}.  

\begin{figure*}
    \centering
    \includegraphics[width=0.9\linewidth]{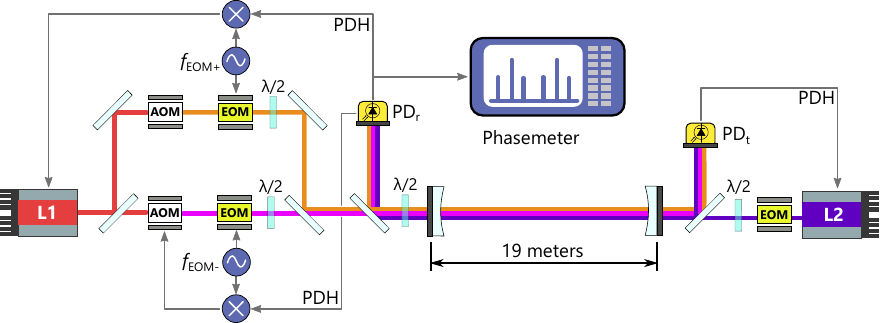}
    \caption{The optical system for the prototype. The fields $E_{\parallel_-}$ (shown in orange) and  $E_{\parallel_+}$ (shown in pink) are generated by frequency shifting laser L1 using \acp{AOM} in a double pass configuration (not shown). The field $E_{\perp_0}$ (shown in purple) is provided by laser L2. All three fields are frequency stabilized to a different resonance of a 19\,m optical cavity. $\rm PD_r$ is used to sense the interference beatnotes between the fields and this signal is then sent to a phasemeter to track the frequencies of the beatnotes.}
    \label{fig:Proto}
\end{figure*}

A prototype setup was built to test the technique on a 19\,m cavity. A diagram of the experiment can be seen in Figure~\ref{fig:Proto}. The three laser fields were generated with two lasers operating with a wavelength of 1064\,nm, one of which, L1 in the diagram, is split into two paths. Both paths from L1 are frequency shifted by their own \ac{AOM} in a double pass configuration (not shown in Figure~\ref{fig:Proto}). 
An \ac{EOM} in each path also generates phase modulation sidebands such that each field can be individually stabilized to a cavity resonance using the \ac{PDH} technique. A single \ac{PD} in reflection of the cavity (labeled $\rm PD_r$ in Figure~\ref{fig:Proto}) is used as the sensor in both loops. The feedback on one of the paths is sent directly to the frequency of L1 (the field emerging from this path will be referred to as $E_{\parallel_+}$ and is shown as the orange line in Figure~\ref{fig:Proto}), while in the other path the \ac{AOM} is used as a frequency actuator (this field will be referred to as $E_{\parallel_-}$ and is shown as pink line in Figure~\ref{fig:Proto}). 
The two paths each have a \ac{HWP} and are combined at a power beam splitter to allow any configuration of linear polarization states to be used. Faraday isolators, not shown in Figure~\ref{fig:Proto} for simplicity, are used to prevent light from being reflected back to the lasers.

A second laser (L2) on the opposite side of the cavity provides the third field which will be referred to as $E_{\perp_0}$. 
This laser also has an \ac{EOM} in its beam path such that it can be \ac{PDH} stabilized to a cavity resonance using $\rm PD_t$ as a sensor. 
$\rm PD_r$ on the L1 table is then used as the primary sensor of the interference beatnotes between the three fields. 


%

Since non-polarizing beam splitters are used to inject all fields to the cavity, the setup does not restrict the polarization states of the fields. This allows for null measurements with all fields in the same polarization state to test the sensitivity of the setup independent of polarization effects. However, when the setup is configured to measure the dynamic birefringence of the cavity, $E_{\perp_0}$ is aligned to an orthogonal polarization axis from $E_{\parallel_+}$ and $E_{\parallel_-}$. When this is the case, the interference beatnotes are generated by orienting the wave-plate before $\rm PD_r$ to rotate all polarization states by $45^\circ$ with respect to the polarizing beamsplitter  just before $\rm PD_r$. Therefore all fields are split equally and the interference beatnotes can be measured at either port of the polarizing beamsplitter.



\subsection{Frequency Read-out}

\begin{figure}[]
    \centering
    \includegraphics[width=0.9\linewidth]{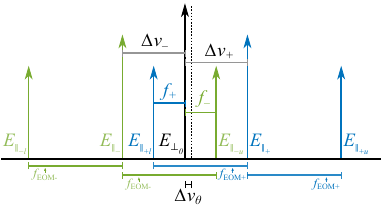}
    \caption{The frequency configuration of the fields present at $\rm PD_r$ in the prototype setup. To sense the changes in $\Delta\nu_-$ and $\Delta\nu_+$, the interference beatnotes at $f_+$ and $f_-$ where measured.}
    \label{fig:freq_proto}
\end{figure}
It can be difficult to measure the direct interference beatnotes $\Delta \nu_+$ and $\Delta \nu_-$ because they are at nearly the same frequency. However, the interference beatnotes between $E_{\perp_0}$ and the phase modulation sidebands on $E_{\parallel_+}$ and $E_{\parallel_-}$, shown in  
Figure~\ref{fig:freq_proto}, can be used instead. 
The sidebands track the frequency changes of their carrier fields. Therefore, any changes in the resonance frequencies of the cavity will be transferred to the sidebands. Measurements of the clock jitter in the oscillators were performed, and the noise projected from these results was well below the expected cavity birefringence noise.

The phase modulation frequency of  $f_{_{\rm EOM+}}$ is $\simeq11.8$\,MHz. Therefore, the $E_{\parallel_+}$ lower sideband ($E_{\parallel_{+l}}$ in Figure \ref{fig:freq_proto}) resides at a lower frequency than $E_{\perp_0}$ because  $f_{_{\rm EOM+}}$ is larger than the cavity \ac{FSR}. 
The actual beatnote frequency used to monitor $\Delta\nu_+$ was thus 
\begin{equation}
f_+=\rm11.8000094\,MHz-(\overbrace{f_0-\Delta\nu_\theta}^{\Delta\nu_+}),    
\end{equation}
or 3.9082\,MHz. While the values of the beatnote frequencies are know with sub-Hz accuracy, fluctuations in the cavity length, and hence \ac{FSR} ($f_0$), cause them to change over the course of data taking and therefore they are only given with a precision of 100\,Hz here. Nevertheless, these changes are correlated, so the differences between the frequencies can be given with sub-Hz precision.

To monitor changes in $\Delta \nu_-$ the $E_{\parallel_-}$ upper sideband ($E_{\parallel_{-u}}$ in Figure \ref{fig:freq_proto}) was used with a modulation frequency of $\simeq12.4$\,MHz. The frequency of the beatnote used to monitor $\Delta\nu_-$ was therefore \begin{equation}
f_-=\rm12.3999815\,MHz-(\overbrace{f_0+\Delta\nu_\theta}^{\Delta\nu_-}),    
\end{equation} 
or 4.5081\,MHz. 

The phasemeter measuring the frequencies of the beatnotes operated with a bandwidth of 10\,kHz. The initial sampling rate of the analog-to-digital converters was $10^9$ samples per second with the phasemeter data then being down sampled to 596\,Hz after the data was filtered by a finite-impulse-response filter. The data stream from the phasemeter consisted of a continuous data set of the frequency, phase, and amplitude of the beatnote it was measuring.


\subsection{Optical cavity}
\label{Sec:cav}
A detailed characterization of the optical cavity used in this study is available in Ref.~\cite{Spector:24}. The cavity is formed by two 2" diameter mirrors with a single pass length of $(18.9938\pm0.0001)$\,m leading to an \ac{FSR} of $(7.89186\pm0.00003)$\,MHz. The cavity mirrors are mounted directly to separate optical tables and are contained within a single vacuum system. The lasers are injected via vacuum windows at both ends of the cavity. The nominal storage time is $(1.30\pm0.1)$\,ms, corresponding to a finesse of $32,200\pm200$. The input mirror of the cavity has a transmissivity of $(90.0\pm0.5)$\,ppm, while the output mirror transmissivity is $(89.8\pm0.4)$\,ppm. The excess optical losses in addition to the mirror transmissivities were found to be 15\,ppm. The power buildup of the cavity is 9470 and with 20\,mW typically injected for a measurement the cavity circulating power reaches nearly 200\,W. Typically values for the field overlap between the injected lasers and the cavity eigenmode were measured to be between 92\% and 97\%. The geometry of the cavity is nearly confocal, with an average radius of curvature of the mirrors measured via the higher order mode spacing to be 19.95\,m. 

The static difference in the cavity resonance frequencies due to birefringence, $\Delta\nu_{\theta}$, can be calculated from the measured frequencies using the expression:
\begin{align}
\label{Eq:del_nu_3f}
\Delta\nu_{\theta}&=\frac{\Delta\nu_+ - \Delta\nu_-}{2}\\
&=f_+-f_{\rm EOM +}-(f_- - f_{\rm EOM-}) .  
\end{align}
This gives $\Delta\nu_{\theta}=4.1\pm0.3$\,Hz. Section~\ref{Sec:HWP_rot} describes how the static birefringence of the cavity was also measured by observing the change in the beatnote frequency of the fields stabilized to the cavity resonances while rotating the input polarization states using a \ac{HWP}. This technique resulted in a measured value of $\Delta\nu_{\theta}=4.25\pm0.02$\,Hz. The agreement between these methods is clear.

According to the measurement with the rotating polarization states, the static birefringence of the cavity is $(0.538\pm 0.003)\times10^{-6}$, leading to $\Delta \theta=(3.38\pm 0.02)\times10^{-6}$\,rad. The major and minor polarization axes of the cavity are aligned at $40^\circ$ with respect to the vertical and horizontal directions and the polarization states of the input fields are aligned to them. In the future, we plan to rotate the mirrors so that the fast axis of the cavity is aligned to the vertical direction, but this has not yet been attempted in this setup.

\section{Results}
\label{Sec:Results}

\begin{figure}
    \centering
 \subfloat[ Time series of the beatnote frequency drift 
    \label{fig:ts}]{
    \includegraphics[width=0.9\linewidth]{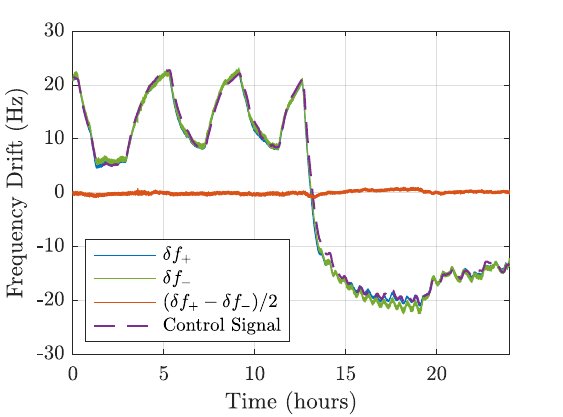}
    }
    
        \subfloat[ Allan deviation of the beatnote frequency drift \label{fig:adev}]{
        \includegraphics[width=0.9\linewidth]{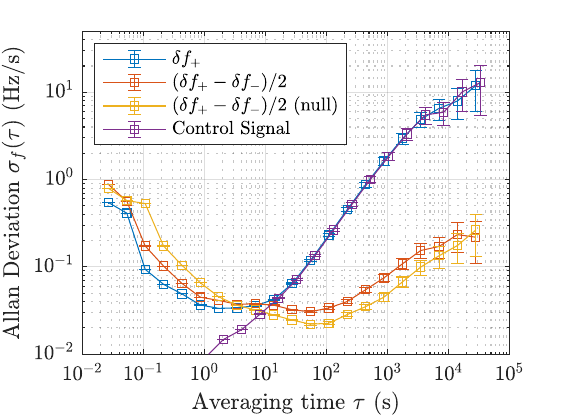}
        }
 
    \subfloat[ \ac{ASD} of the measurement noise in the setup
    \label{fig:sens}]{
    \includegraphics[width=0.9\linewidth]{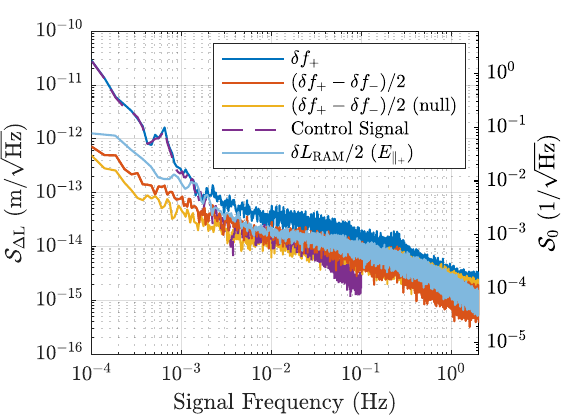}
    }
    \caption{Shown above are a time series (a), Allan deviation (b), and \ac{ASD} (c) of the measurement of the frequency drifts of the interference beatnotes. Here the $\delta f_+$ is shown in blue, $\delta f_-$ is shown in green, and the measurement of $(\delta f_+-\delta f_-)/2$ is shown in orange. A separate null measurement of $(\delta f_+-\delta f_-)/2$ is shown in yellow, with the three fields configured to have the same polarization. The calibrated feedback control signal sent to the laser temperature is shown in purple. A projection of the out-of-loop noise induced by residual amplitude modulation in the $E_{\parallel_+}$ \ac{PDH}  frequency stabilization loop is shown in light blue. \label{fig:noise}}
\end{figure}

A time series of a 20\,hour measurement of the beatnote frequencies measured by the prototype setup is shown in Figure~\ref{fig:ts}. A moving average filter with a time constant of 0.67\,s is applied to the data in this plot. The frequency drifts of the beatnotes $\delta f_-$, $\delta f_+$, and their superposition from Equation~\ref{Eq:sup_beat} $(\delta f_+-\delta f_-)/2$ are shown as the green, blue, and orange traces of this plot respectively. The purple trace gives the temperature control signal sent to the laser crystal, calibrated in terms of the frequency changes of the cavity \ac{FSR}. While the frequency drift of the laser contributes to the control signal, it is primarily driven by the length changes of the cavity. From this plot it is apparent that the length drift of the cavity is common to both $\delta f_+$ and $\delta f_-$, but this can be canceled using the superposition $(\delta f_+-\delta f_-)/2$. This superposition reduced the standard deviation in the frequency drift from 15.5\,Hz for $\delta f_+$ and 15.9\,Hz for $\delta f_-$ to 0.3\,Hz for $(\delta f_+-\delta f_-)/2$. The standard deviation of the calibrated control signal from its mean value over this time was 15.6\,Hz.

The Allan deviation of the frequency drifts from this measurement is shown in Figure~\ref{fig:adev} with error bars that give the $1\,\sigma$ statistical uncertainty in the data. Here, an additional null measurement is shown in yellow, in which all three fields stabilized to the cavity resonances were in the same polarization state. This measurement was performed to assess whether the measurements were limited by the control system, specifically the precision of the \ac{PDH} loops, versus polarization effects in the setup. The Allan deviation shows that both the normal and null measurement of $(\delta f_+-\delta f_-)/2$ demonstrate significantly better stability for averaging times longer than 20\,s, compared to the measurement of the frequency drift of $\delta f_+$. The stability of $\delta f_+$ for these times is limited by the cavity length noise as shown by its agreement with the control signal. In addition to this, the plot shows good agreement between the normal and null measurements of $(\delta f_+-\delta f_-)/2$, out to averaging times of 30,000\,s. 

The \ac{ASD} of the noise achieved in each of the measurements is shown in Figure~\ref{fig:sens}.
Here, the $x$-axis gives the Fourier frequency of the noise, while the $y$-axis to the left gives the sensitivity of the measurement in terms of the cavity's differential single-pass optical path length ($\mathcal S_{\Delta \rm L}$). The $y$-axis on the right shows the sensitivity relative to the \ac{FWHM} of the cavity ($\mathcal S_{0}$). 
The noise in the normal and null measurements of $(\delta f_+-\delta f_-)/2$ show good agreement across the full range of frequencies. As expected, the length noise of the cavity leads to an increase in the noise of $\delta f_+$ below 2\,mHz. The fact that the \ac{ASD} of the noise in the measurements of $(\delta f_+-\delta f_-)/2$ is below the noise measured in $\delta f_+$ at such frequencies is a critical result. This demonstrates that the prototype setup is capable of sensitivities below the noise floor established by the fluctuations in the length of the cavity.



In the measurements of $(\delta f_+-\delta f_-)/2$, an \ac{ASD} of $4\times10^{-14}\rm\,m/\sqrt{Hz}$ (or 0.2\%/$\sqrt{\rm Hz}$ of the \ac{FWHM} of the cavity) was achieved at a frequency of 3\,mHz. The current noise floor of $(\delta f_+-\delta f_-)/2$ is also roughly a factor of 2 better than the measurements of $\delta f_+$  above 2\,mHz. 

The results shown in Figure~\ref{fig:sens} are representative of the sensitivity of the system in an optimally aligned state. Maintaining such an alignment over the course of a \ac{VMB} search lasting millions of seconds would require automated alignment control or periodic pauses in the data taking to realign the system. This is possible, as the cross-correlation of the expected signal waveform with the measured data would not require continuous sampling.

Both the birefringence and null measurements of $(\delta f_+-\delta f_-)/2$ are believed to be currently limited by out-of-loop noise in the \ac{PDH} frequency stabilization systems that is induced by the \ac{RAM} of the lasers at the \ac{EOM} phase-modulation frequencies, as discussed in Section~\ref{Sec:OOL}. 

In the future, we expect that additional noise sources in the prototype setup will have to be addressed. 
Section~\ref{Sec:Cav_Ali} describes a measurement of the alignment noise of the cavity mirrors and projects it in terms of  sensitivity limits on $\mathcal S_{\Delta \rm L}$. A discussion of several noise sources which are not expected to limit the sensitivity of the experiment, including laser shot noise and polarization noise, along with the dispersion effects in the mirror coatings, can be found in Appendices~\ref{App:Shot}, \ref{App:Pol}, and \ref{App:Dis} respectively. 

Finally, Section~\ref{Sec:HWP_rot} reports on an independent measurement of the cavity birefringence using a rotating \ac{HWP}.

\subsection{RAM Induced Out-of-Loop Noise}
\label{Sec:OOL}

The phase modulation of the laser at the \ac{EOM} frequency will convert to amplitude modulation in reflection of the cavity. This is due to a phase shift in the carrier field with respect to the phase modulation sidebands. When the laser frequency is close to the resonance of the optical cavity, this amplitude modulation is used to generate the \ac{PDH} error signal. There are a number of effects that can also cause \ac{RAM} at the \ac{EOM} driving frequency, such as etalons forming in the optics and polarization modulation introduced by the \ac{EOM}. \ac{RAM} can cause problems in the \ac{PDH} loops as it can lead to offsets at the error point \cite{Kedar:24, Gillot:22}. If the \ac{RAM} changes at all, this will cause error point noise that the feedback loop does not suppress. An example of this is shown in Figure~\ref{fig:sens} as the light blue trace. Here, the noise at the error point of the $E_{\parallel_+}$ \ac{PDH} control loop was measured with the cavity blocked. This noise was then calibrated in terms of the cavity length $\delta L_{_{\rm RAM}}$, using
 \begin{equation}
\delta L_{_{\rm RAM}}(t) \simeq \frac{\lambda}{2f_0}\left(1-\eta^2\frac{2T_i}{A}\right)\frac{V(t)}{m_{_{\rm ES}}}.
\label{Eq:OOL}
\end{equation}
Here, $V(t)$ is the time series of the voltage noise measured at the error point of the \ac{PDH} lock with the cavity blocked, while  $m_{_{\rm ES}}$ is the slope of the \ac{PDH} error signal in V/Hz (details on this measurement can be found in Appendix~\ref{App:ERR}). The spatial overlap between the cavity and the laser is given by $\eta$, $T_i$ is the power transmissivity of the input coupling mirror, and $A$ is the total optical losses that the circulating field experiences during one round trip through the cavity including the mirror transmissivities, as well as scattering and absorption in the reflective coatings of the mirrors. In this case, the term $\frac{2T_i}{A}$ gives the field reflection coefficient of the cavity on resonance. A derivation of Equation~\ref{Eq:OOL} and a discussion of the assumptions it relies on are given in Appendix~\ref{App:RAM}. 

Based on Figure~\ref{fig:sens} it appears as though the \ac{RAM} in $E_{\parallel_+}$ (the light blue trace) may be responsible for the excess noise seen in the measurements of $(\delta f_+-\delta f_-)/2$ (the orange and yellow traces). The out-of-loop noise was also measured in the frequency control loops for $E_{\perp_0}$ and $E_{\parallel_-}$, but the projected noise is significantly lower than for the $E_{\parallel_+}$ control loop. A plot of the projections of these measurements can be found in Appendix~\ref{App:RAM_other} for all three fields. Therefore, mitigating the noise in the $E_{\parallel_+}$ loop may significantly improve the sensitivity of the setup as it appears to set the current noise floor.

\subsection{Cavity Alignment Noise}
\label{Sec:Cav_Ali}

The alignment noise of the cavity mirrors is a potential source of birefringence noise in this measurement scheme. The position of the cavity eigenmode on the mirrors is determined by their relative alignment. Since the mirror's differential reflected phase for orthogonal polarization states ($\Delta\theta$ in Equation~\ref{Eq:Del_nu}) is spatially dependent, the cavity alignment noise can induce birefringence noise.

\begin{figure}
    \centering
 \subfloat[Cavity eigenmode position drift measurement
    \label{fig:point_plot}]{
    \includegraphics[width=0.9\linewidth]{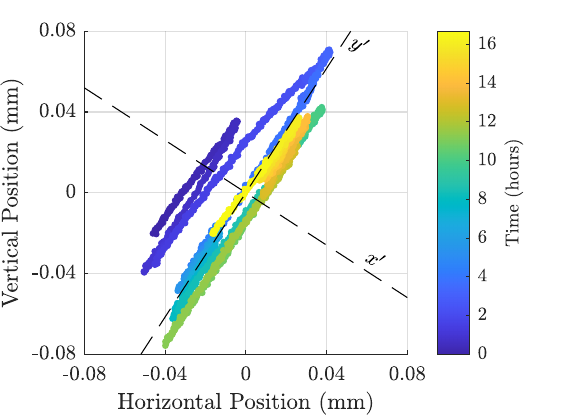}
    }

\subfloat[Projected noise \ac{ASD} due to position drift
    \label{fig:point_psd}]{        
        \includegraphics[width=0.9\linewidth]{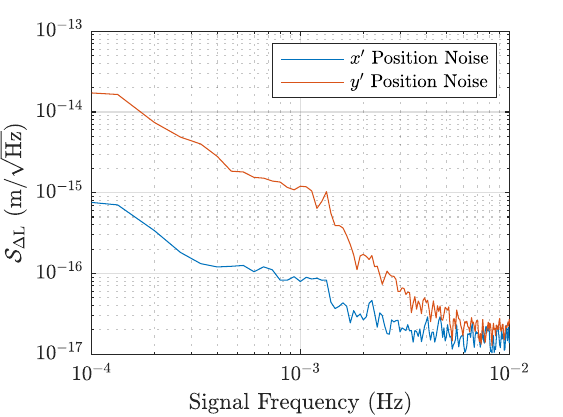}}
    \caption{The results of a 16 hour measurement of the eigenmode position in transmission of the cavity is shown in (a). The position of the transmitted beam was measured by a camera placed in transmission and is normalized by the beams $1/e^2$ radius of 3\,mm. As the plot shows, most of the drift occurs along the $y'$ axis indicated by the dashed line. \acp{ASD} of the position drift along the $x'$ and $y'$ axes were calculated and projected in terms of $\mathcal S_{\Delta \rm L}$ in (b). \label{fig:point}}
\end{figure}

A 16 hour measurement of the drift in the position of the cavity eigenmode is shown in Figure~\ref{fig:point_plot}. Here, $E_{\parallel_+}$ was frequency stabilized to the cavity ($E_{\parallel_-}$ and $E_{\perp_0}$ were switched off) and a camera  placed in transmission recorded the position of the beam. The RMS deviations from the mean were 24\,{\textmu}m and 34\,{\textmu}m for the horizontal and vertical positions respectively. 

The transmitted beam also appeared to move predominantly on the axis labeled $y'$ indicated by the black dashed line. On this axis the RMS deviations from the mean position were 38\,{\textmu}m. 
The orthogonal axis, labeled as $x'$, showed RMS deviations of 9\,{\textmu}m.

This motion of the beam spot on the mirror can be projected in terms of birefringence noise using the spatial distribution of the cavity mirror birefringence. Although we did not measure mirror birefringence maps in this study, we projected the birefringence noise induced by this motion using
 \begin{equation}
\delta L_{_u}(t) \simeq \delta u\frac{\lambda}{2}\frac{d\,\Delta\theta}{du}.
\label{Eq:xy_S}
\end{equation} In this equation $\delta u$ refers to the position noise of the beam spot on an arbitrary axis $\hat u$ and $\frac{d\,\Delta\theta}{du}$ is the spatial derivative of the birefringence on that axis. The projections rely on the assumption that $\Delta\theta$ varies by 20\% between independent points~\cite{micossi1993measurement}. As we assume the cavity birefringence is spatially coherent over a beam diameter, the 6\,mm $1/e^2$ beam diameter leads to an expected spatial derivative in the birefringence of $1.8\times10^{-8}\rm/mm$.

The results of these projections in terms of the limit on the sensitivity of the prototype setup are shown in Figure~\ref{fig:point_psd}. The blue trace refers to the projections of the position noise along the $x'$ axis (from Figure~\ref{fig:point_plot}), while the orange trace refers to the noise on the $y'$ axis. The pointing measurements themselves were limited by a sensitivity of 2\,{\textmu}m/$\sqrt{\rm Hz}$ which projects to a noise floor of $1.8\times10^{-17}\rm\,m/\sqrt{\rm Hz}$ in this plot. 

These results indicate that an automatic alignment system may be needed to suppress the pointing induced birefringence noise below the prototype's sensitivity goal of $10^{-17}\rm \,m/\sqrt{\rm Hz}$. 
However, the projections are encouraging. At 3\,mHz the noise along the $y'$ axis would only need to be suppressed by a factor of 6 to reach the sensitivity goal.

\subsection{Cavity Birefringence Measurement}
\label{Sec:HWP_rot}

The static birefringence of the cavity was measured by stabilizing two orthogonally polarized fields to adjacent cavity resonances, continuously rotating their polarization states at the input of the cavity, and measuring their beatnote frequency. The fields were both generated by frequency shifting L1 using \ac{AOM}s. They were then independently stabilized to resonances of the cavity, while their polarization states were modulated by rotating a \ac{HWP} directly before the cavity. A \ac{PD} in transmission of the cavity measured the interference beatnote and a phasemeter was used to track its frequency. The cavity birefringence was determined from the amplitude of the changes induced in the beatnote frequency. As the fields are aligned to orthogonal polarization states, rotating the \ac{HWP} will induce a modulation in the beatnote frequency with a peak-to-peak amplitude $2\Delta\nu_{_\theta}$ at four times the \ac{HWP} rotation frequency $f_{\rm rot}$  \cite{hall2000measurement}. 

\begin{figure}
    \centering
 \subfloat[ Beatnote frequency while rotating the \ac{HWP}
    \label{fig:rot}]{
    \includegraphics[width=0.9\linewidth]{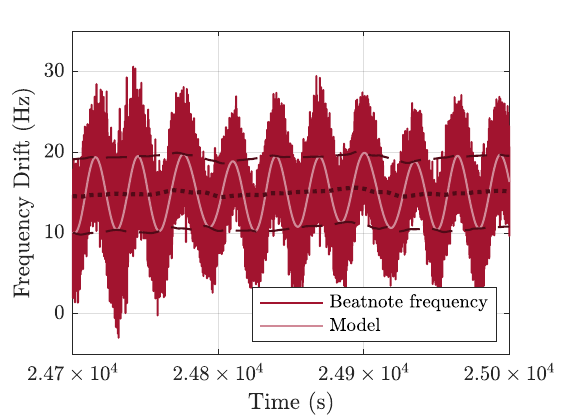}
    }

\subfloat[Time series of the amplitude of the birefringence 
    \label{fig:amp}]{
        \includegraphics[width=0.9\linewidth]{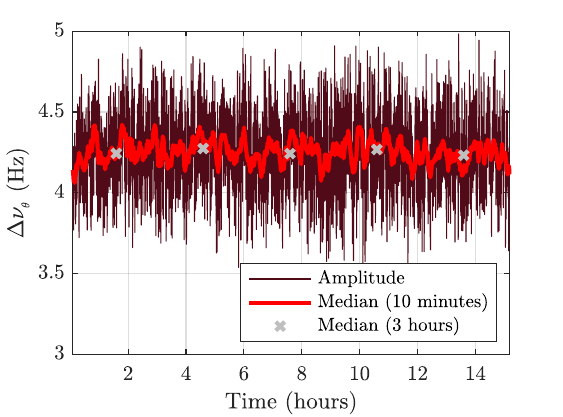}
        }
 
\subfloat[\ac{ASD} of the amplitude of the birefringence
    \label{fig:amp_lsd}]{        
        \includegraphics[width=0.9\linewidth]{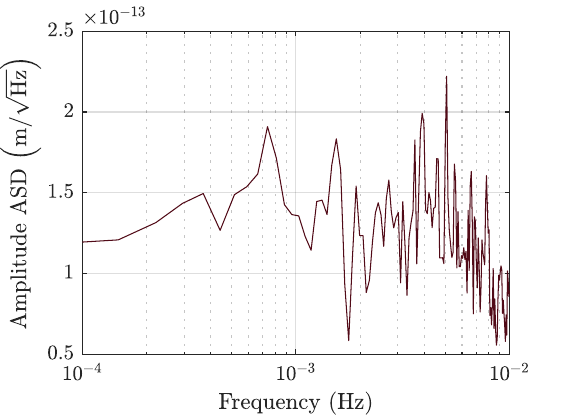}}
    \caption{The results of measuring the beatnote frequency while rotating a \ac{HWP} in the path of the fields injected to the cavity over the course of nearly 15 hours are shown above. The cavity birefringence leads to the oscillations seen in the section of the time series shown in (a). The result of fitting a sinusoidal model to the time series of the beatnote frequency is shown as the pink trace in (a). The time series of the amplitude  calculated from the fit of the model is shown in (b), along with the median amplitude calculated over 10\,minute and 3\,hour sections of data. The \ac{ASD} of the changes in the amplitude is shown in (c). \label{fig:HWP}}
\end{figure}

The dark red trace in Figure~\ref{fig:rot} shows a 300 second section of a nearly 15 hour measurement of the beatnote frequency while rotating the \ac{HWP}. 
To find the amplitude of the oscillations in the beatnote frequency, the following model,
\begin{equation}
    \Delta\nu_{\theta}\,\sin{\left(8\pi f_{\rm rot} (t'-t_0)  + \phi_{\rm rot} \right)}  + \delta \nu_{L,0}+  \frac{d\, \delta \nu_L}{dt} \,(t'-t_0),
\end{equation}
was fit to the sections of the data spanning 50 seconds, centered on $t_0$. Thus for every time $t$ in the data, a set of the parameters in the equation could be produced. Here, $\phi_{\rm rot}$ defines the phase of the \ac{HWP} rotation, with $\delta \nu_{L,0}$ representing a static offset in the beatnote frequency and $\frac{d\, \delta \nu_L}{dt}$ representing the linear drifts in this offset. These offsets are related to the beatnote frequency changing due to changes in the cavity \ac{FSR}. 

The pink curve in Figure~\ref{fig:rot} shows the result of the fitted model. Here the dotted line gives $\delta \nu_{L,0}$ while the dashed lines show the maximum and minimum of the oscillations. The birefringence signal $\Delta\nu_{\theta}$ over the full 15\,hours is extracted from the fit and shown as a function of time in Figure~\ref{fig:amp}. The darkest line shows the raw time series of the fitted values, while the red line shows the median of the time series calculated over 10 minute long sections of data, and the gray x's show the median amplitude calculated over three hour sections of the data. The median of $\Delta\nu_{\theta}$ over the entire run was $4.25\pm0.02\,\rm Hz$. The uncertainty represents the root-mean-square drifts of the median of $\Delta\nu_{\theta}$ over long time periods ($>\rm 3~hours$). 
This corresponds to a cavity birefringence of $(538\pm 3)\times10^{-9}$ and demonstrates that the technique is capable of measuring the static birefringence of the cavity to a precision of several ppb.


Figure~\ref{fig:amp_lsd} shows the result of converting $\Delta\nu_{\theta}$ to  $\Delta L_{\theta}$ using the relation $\frac{\Delta \nu}{\nu} = -\frac{\Delta L}{L}$ and then taking the \ac{ASD} of the fluctuations in $\Delta L_{\theta}$. This gave a level of roughly  $1.3\times10^{-13}\rm\,m/{\sqrt{Hz}}$ at 3\,mHz, which is also the mean value of this spectrum below 7\,mHz. Here, the noise floor is determined by the \ac{ASD} of the measurement noise at $\sim30$\,mHz. The sensitivity achieved in Figure~\ref{fig:amp_lsd} is roughly a factor of 5 higher than the \ac{ASD} of $\delta f_+$ at 30\,mHz in Figure~\ref{fig:sens} due to the fact that the rotation of the \ac{HWP} introduced additional frequency noise. The reason for this is under investigation.

Nevertheless, it is worthwhile to note that at frequencies close to 100\,{\textmu}Hz this measurement had a lower noise level than what was achieved in the measurements of the frequency drift of $\delta f_+$ and $(\delta f_+-\delta f_-)/2$ without rotating the \ac{HWP} (seen in Figure~\ref{fig:sens}). This shows that rotating the polarization states and observing the oscillations of the beatnote frequencies can lead to a better noise performance at low frequencies than in the baseline design. This is a result of the measurement noise being up-converted to higher frequencies (at four times the \ac{HWP} rotation frequency). Effects coming from inside the cavity, such as \ac{VMB} or the intrinsic cavity birefringence noise,  will still couple to the measurement at the magnet modulation frequency.

\section{Conclusions}
\label{Sec:Conc}

This work represents the first demonstration of a measurement of the birefringence of an optical cavity by measuring the changes in three different resonances as was first described in Ref. \cite{hall2000measurement}. This experiment was performed on a 19\,m optical cavity, meaning that the setup is already large enough to potentially incorporate a superconducting dipole magnet like those used in particle accelerators. 

The ultimate goal of the experiment is to demonstrate a sensitivity sufficient to test the \ac{QED} prediction of the \ac{VMB} effect using the string of 24 superconducting dipole magnets assembled for the ALPS\,II experiment. While a measurement of \ac{VMB} at the predicted amplitude would be a first macroscopic confirmation of nonlinear \ac{QED}, measuring a deviation from this prediction would be even more exciting as it would be evidence of new physics. 

According to the \ac{QED} prediction, the ALPS\,II magnet string is capable of producing a \ac{VMB} signal with an amplitude of $2.37\times10^{-20}$\,m in terms of the differential single pass length for orthogonal polarization states. The sensing scheme would therefore require a differential length sensitivity, at the characteristic frequency of the modulation of the magnetic field, on the order of $10^{-17}\rm\,m/\sqrt{Hz}$. This is equivalent to a fractional sensitivity in terms of the \ac{FWHM} of the cavity of $2\,\rm ppm\,/\sqrt{Hz}$,
assuming a cavity finesse of 100,000. 
To reach this goal, the sensitivity of the prototype must improve three to four orders of magnitude from its current level in the relevant frequency range. 

There is evidence that the setup is currently limited by \ac{RAM}-induced out-of-loop noise in the laser frequency stabilization systems. This is being addressed with a redesign of the optics on the injection paths to the cavity, including a system that provides active \ac{RAM} suppression. The aim of this effort will be to improve the sensitivity of the prototype to a level capable of measuring the intrinsic birefringence noise of the 19\,m cavity. 
This would be in line with state-of-the-art laser frequency stabilization systems, such as in Ref.~\cite{Kedar:24}, which demonstrated a fractional stability of $1.2\rm\,ppm/\sqrt{\rm Hz}$ of the cavity linewidth at a Fourier frequency of 10\,mHz. This is comparable to what a \ac{VMB} measurement using the ALPS\,II magnet string would require.

A factor that has limited the previous generation of  \ac{VMB} experiments was the intrinsic birefringence noise of the optical cavity. Compared to these experiments, the full scale setup using the ALPS\,II magnet string will be vulnerable to the cavity birefringence noise due to the low magnet modulation frequency. However, it will benefit from the longer cavity that produces larger beam sizes on the mirrors. If the sensitivity of the experiment is limited by this noise, we project that it should be able to test the \ac{QED}-predicted \ac{VMB} amplitude with a signal to noise ratio of 3 after 1,600,000 seconds of integration time. Therefore, the prospects for a measurement of \ac{VMB} using the ALPS\,II magnet string are very promising. 

\section*{Acknowledgments}

We would like to thank Jan P\~old, Michael T. Hartman, Sandy Croatto, David Reuther, and the rest of the ALPS collaboration for helping to develop the infrastructure for the prototype lab and generally supporting this research. We are also grateful to Jacob Egge, Aldo Ejlli, Hartmut Grote, Axel Lindner, Guido Mueller, David B. Tanner, and Benno Willke for many interesting discussions and their feedback regarding the manuscript. Finally, this work would not have been possible without the help of the infrastructure groups at \ac{DESY}.

The work is supported by the Deutsche Forschungsgemeinschaft (DFG, German Research Foundation) under Germany’s Excellence Strategy – EXC 2121 ``Quantum Universe" – 390833306 and - grant number WI 1643/2-1, the Partnership for Innovation, Education and Research (PIER) of DESY and Universit\"at Hamburg under PIER Seed Project - PIF-2022-18, the German Volkswagen Stiftung, the National Science Foundation - grant numbers PHY-2110705 365 and PHY-1802006, the Heising Simons Foundation - grant numbers 2015-154 and 2020-1841, the UK Science and Technologies Facilities Council - grant number ST/T006331/1.

\bibliography{VMB.bib} 

\appendix


\section{Noise Projections}
In the following Appendices~\ref{App:Shot}, \ref{App:Pol}, and \ref{App:Dis}, projections of the measurement noise due to laser shot noise, polarization noise, and dispersion in the mirror coatings are calculated, respectively.

\subsection{Laser Shot Noise}
\label{App:Shot}
With 1\,mW of power incident on the cavity from each of the fields, the shot noise limit of the \ac{PDH} sensing is projected to be $6\times10^{-19}\rm\,m/\sqrt{Hz}$ \cite{black2001introduction}. 
Adding the shot noise quadratically for the superposition of the beatnotes results in a combined shot noise of less than $10^{-18}\rm\,m/\sqrt{Hz}$, well below the noise measured in the prototype and the sensitivity goal for the full scale experiment.

\subsection{Laser polarization noise}
\label{App:Pol}
The stability of the polarization states of the fields at the input of the cavity was considered. 
Long term measurements of these fields using a diagnostic polarimeter were performed at a position just before the cavity mirror as well as in transmission of the cavity. These suggested no evidence of polarization noise in the laser fields to within the 4\,mrad sensitivity of the device. Therefore, differential length noise induced by changes in the polarization states of the fields at the cavity input can be excluded above $4\times 10^{-18}\rm\, m/\sqrt{Hz}$.


\subsection{Dispersion in the mirror coatings}
\label{App:Dis}
The effects of dispersion in the dielectric coatings of the cavity mirrors were also calculated. The length noise of the cavity could couple to the measurement if these coatings show strong enough dispersion, as each of the three fields would experience a different optical path length through the cavity and a frequency-dependent free spectral range. With a group delay dispersion in the reflectance of the optical coatings expected to be on the order of -8\,$\rm fs^2$ (data from manufacturer) and a frequency spacing between the beatnotes of 8\,MHz, this would correspond to a differential \ac{FSR} between the upper and lower sidebands of only 0.3\,{\textmu}Hz. This projects to an equivalent length noise in the full scale experiment of $3\times 10^{-25}{\rm\,m/\sqrt{Hz}}$.

\section{Coupling of RAM to PDH loops}
\label{App:RAM}
 
To calculate the impact of the \ac{RAM} on the error signal offset, first a field with both phase and amplitude modulation should be considered:
\begin{equation}
    E = E_0 e^{i(\omega t + a_\phi \sin\Omega t) + a_0 \sin\Omega t}.
\end{equation}
Here $\omega$ is the angular frequency of the carrier field, while $\Omega$ is the angular frequency driving the \ac{EOM}. The modulation depth of the phase modulation is given by $a_\phi$, while the modulation depth of the amplitude modulation is given by $a_0$. In both cases we assume $a\ll1$.
The power in this field measured by a \ac{PD} can then be expressed as 
\begin{equation}
    P \simeq P_0 + 2P_0a_0 \sin\Omega t,
\label{APP_EQ:Pow}
\end{equation}
using the small angle approximation and ignoring higher order terms in $a$. In this equation, the power in the carrier is given by $P_0$. 

When considering how the field interacts with the cavity, it is convenient to express it as a carrier with real and imaginary sidebands due, respectively, to the amplitude and phase modulation:
\begin{equation}
\begin{aligned}
      E \simeq E_0 e^{i\omega t} \Big[ 1 &+ i\frac{a_\phi}{2} \left(e^{i\Omega t} -e^{-i\Omega t}  \right) \\ &+ \frac{a_0}{2} \left(e^{i\Omega t} -e^{-i\Omega t}  \right) \Big]  .
\end{aligned}
\end{equation}
If this field is then incident on a cavity and the carrier is resonant, the first term in the brackets will experience the cavity reflectivity on resonance, while the sidebands, which are assumed to be well off the cavity resonance, will not. Therefore, the field in reflection of the cavity can be expressed as

\begin{equation}
\begin{aligned}
      E \simeq E_0 e^{i\omega t} \Big[ \mathcal{R}(\omega) & + i\frac{a_\phi}{2} \left(e^{i\Omega t} -e^{-i\Omega t}  \right) \\ & + \frac{a_0}{2} \left(e^{i\Omega t} -e^{-i\Omega t}  \right) \Big],    
\end{aligned}
\end{equation}
without considering spatial overlap between this field and the cavity eigenmode. The cavity reflectivity can be represented by \cite{Spector:24}
\begin{equation}
    \mathcal{R}(\omega)\simeq 1-\frac{T_i}{\frac{A}{2}-i\frac{\omega-\omega_0}{f_0}}.
\end{equation}
Here $\omega_0$ refers to the angular frequency of the nearest cavity resonance.

Assuming a small frequency offset between the field and the resonance of the cavity such that $\Delta\omega=\omega-\omega_0\ll\frac{f_0A}{2}$, the previous expression for $\mathcal{R}(\omega)$ can be simplified to 
\begin{equation}
    \mathcal{R}(\omega)\simeq 1-\frac{2T_i}{A} + i\frac{4T_i}{A^2 f_0}\Delta\omega.
\end{equation}
Using the expression above, the power in reflection of the cavity can be calculated with 
\begin{equation}
\begin{aligned}
      P \simeq  P_0 \bigg[ \left( 1-\frac{2T_i}{A} \right)^2  &+  \frac{8T_i a_\phi}{A^2 f_0}\Delta\omega \sin \Omega t \\ &+2a_0\left( 1-\frac{2T_i}{A} \right) \sin \Omega t\bigg]  .
\end{aligned}
\end{equation}
To generate the error signal the power measured at the \ac{PD} is then demodulated at the angular frequency $\Omega$. The amplitude of the error signal in voltage is then
\begin{equation}
V_{\rm ES} = \frac{m_{\rm ES}}{2\pi}\Delta\omega  +2a_0 V_0\left( 1-\frac{2T_i}{A} \right).
\label{APP_EQ:ES_prop}
\end{equation}
In this equation $m_{\rm ES}$ refers the slope of the error signal in V/Hz, due to the phase modulation, and was substituted for the term $\frac{8T_i a_\phi}{A^2 f_0}P_0$. The second term on the right side of this equation is the error signal offset induced by \ac{RAM} and will be defined as to as $\Delta V_{\rm ES}$:
\begin{equation}
\Delta V_{\rm ES} = 2a_0 V_0\left( 1-\frac{2T_i}{A} \right).
\label{APP_EQ:ES_prop_delta_V_ES}
\end{equation}
Here $V_0$ represents the DC voltage output by the \ac{PD} for the power level $P_0$. It is apparent in this equation, that $\Delta V_{\rm ES}$ is proportional to the field reflectivity of the cavity $\left( 1-\frac{2T_i}{A} \right)$ and goes to zero in the case of a perfectly impedance matching ($A=2T_i$).

The \ac{RAM} in the input field from  Equation~\ref{APP_EQ:Pow}, will be defined as 
\begin{equation}
\Delta V_{\rm ES}' = 2a_0V_0 .
\end{equation}
Therefore, to scale the error signal offset in terms of the input field's \ac{RAM}, the relation
\begin{equation}
\Delta V_{\rm ES} = \left( 1-\frac{2T_i}{A} \right)\Delta V_{\rm ES}'
\end{equation}
can be used. To express this in terms of an error between the laser frequency and the cavity resonance, we solve Equation~\ref{APP_EQ:ES_prop} in terms of $\Delta\omega$:
\begin{equation}
\Delta \omega_{_{\rm RAM}} = \frac{2\pi}{m_{\rm ES}}\Delta V_{\rm ES}.
\end{equation}
This can be formulated in terms of single pass cavity length using the expression $\frac{\Delta L}{L} = -\frac{\Delta \omega}{\omega}$, resulting in the  expression 
\begin{equation}
\Delta L_{_{\rm RAM}} \simeq - \frac{\lambda}{2 f_0 }\left( 1-\frac{2T_i}{A} \right)\frac{\Delta V_{\rm ES}'}{m_{\rm ES}}.
\end{equation}
The spatial overlap between the field and the cavity eigenmode, $\eta$, must still be considered. With this in mind, $\Delta V_{\rm ES}'$ can be expressed as 
\begin{equation}
\Delta V_{\rm ES} = \overbrace{\left( 1-\frac{2T_i}{A} \right)\eta^2\Delta V_{\rm ES}' }^{\text{in-mode}}+\overbrace{\left(1-\eta^2\right)\Delta V_{\rm ES}'}^{\text{out-of-mode}} .
\end{equation}
Here, the first term is the in-mode contribution of the \ac{RAM} to the \ac{PDH} error signal offset while the second term is the out-of-mode contribution.

Simplifying the previous equation and expressing it in terms of $\Delta L_{_{\rm RAM}}$ gives the form of Equation~\ref{Eq:OOL} discussed in the text (where $\delta L_{_{\rm RAM}}$ represents fluctuations of the error signal offset):
\begin{equation}
\Delta L_{_{\rm RAM}} \simeq \frac{\lambda}{2 f_0}\left( 1-\eta^2\frac{2T_i}{A} \right)\frac{\Delta V_{\rm ES}'}{ m_{\rm ES}}.
\end{equation}
It is interesting that the first order coupling could be eliminated in the case of an over-coupled cavity ($2T_i>A$) if the spatial overlap could be maintained such that $\eta^2=\frac{A}{2T_i}$.
Practically, that would be very difficult to achieve, but it speaks to the importance of optimizing the spatial overlap and impedance matching to reduce the coupling of the \ac{RAM}. Furthermore, this derivation relies on the assumption that the field producing the amplitude modulation shares the same spatial eigenmode as the carrier field. This is not necessarily the case and a more thorough exploration of this effect would need to consider the spatial overlap of this field and the carrier, separate from the spatial overlap between the carrier and the cavity eigenmode. Nevertheless, the treatment above is believed to be sufficient to assess the impact of the \ac{RAM} at the \ac{EOM} driving frequency on the out-of-loop noise in the \ac{PDH} frequency stabilization loops.

\section{Projected RAM Measurements}
\label{App:RAM_other}

\begin{figure}[t]
    \centering
    \includegraphics[width=\linewidth]{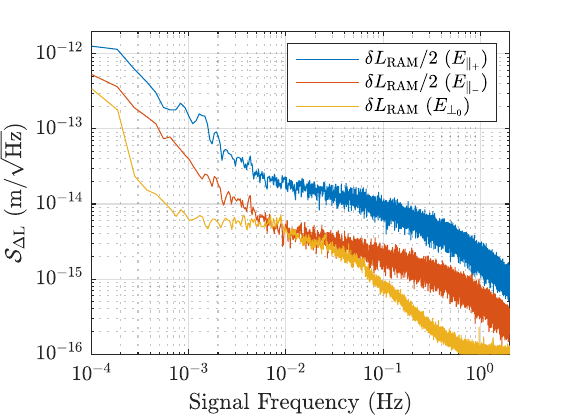}
    \caption{Projections of the noise induced by \ac{RAM} in the \ac{PDH} control loops for $E_{\parallel_+}$, $E_{\parallel_-}$, and $E_{\perp_0}$.}
    \label{fig:RAM_psd}
\end{figure}

Measurements of the error point noise introduced by \ac{RAM} in the \ac{PDH} control loops for $E_{\parallel_+}$, $E_{\parallel_-}$, and $E_{\perp_0}$ are shown in Figure~\ref{fig:RAM_psd} as the blue, orange, and yellow traces respectively. Here, they are projected in terms of their contribution to $(\delta f_+-\delta f_-)/2$ using Equation~\ref{Eq:OOL}. These measurements were performed by blocking the cavity and measuring the changes in the error point of each of the loops. Since the \ac{RAM} of $E_{\parallel_+}$ contributes to $\delta f_+$, and the \ac{RAM} of  $E_{\parallel_-}$ to $\delta f_-$, in both of these projections $\delta L_{\rm RAM}$ is divided by two. As the \ac{RAM} of  $E_{\perp_0}$ introduces an equal and opposite contribution of $\delta L_{\rm RAM}$ to  $\delta f_+$ and $\delta f_-$, its projected contribution is not divided by two. As the plot shows, the \ac{RAM} induced error point noise in the \ac{PDH} lock of $E_{\parallel_+}$ is more than a factor three larger than the same noise in the both of the other loops. 
The reason for this is currently under investigation.

\section{Error Signal Measurement}
\label{App:ERR}

The error signals of the \ac{PDH} loops controlling the frequencies of the fields $E_{\parallel_-}$ and $E_{\parallel_+}$ were directly measured with respect to the difference frequency between the field and the cavity resonance. This was done using a local oscillator field frequency stabilized to a cavity resonance. The field of the error signal being measured was then phase locked to the local oscillator field using an additional photodetector before the cavity (not shown in Figure~\ref{fig:Proto}). The offset frequency of the phase-lock loop was then centered on the \ac{FSR} of the cavity with an additional frequency modulation with a peak-to-peak amplitude of 1\,kHz or roughly 4 cavity line-widths, and a modulation frequency of 1\,Hz. Phase modulation sidebands were generated on the field being tested using its \ac{EOM} and the signal from $\rm PD_r$ was demodulated at the \ac{EOM} driving frequency to generate the error signal. Over the course of a 10 minutes, 1200 error signals were generated and measured with a parallel data stream recording the difference frequency between the fields.

\begin{figure}[b]
    \centering
    \includegraphics[width=\linewidth]{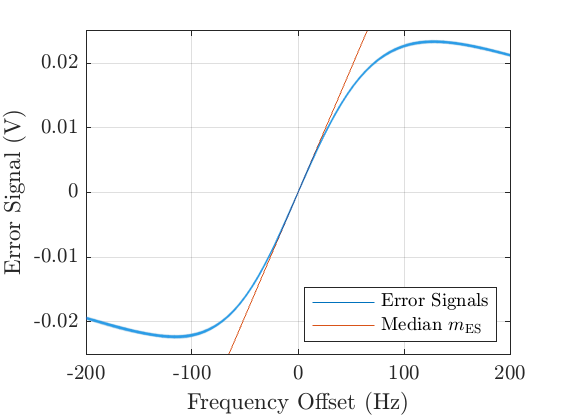}
    \caption{Measurement of the 600 overlaid error signals are shown in blue for the \ac{PDH} loop controlling the frequency of $E_{\parallel_+}$. The horizontal axis gives the frequency offset from the control error point. The orange line shows  the median slope of the error signals obtained at the error point, $m_{\rm ES}$.}
    \label{fig:app_err_sign}
\end{figure}

The results of overlaying 600 of these error signals are shown in Figure~\ref{fig:app_err_sign} as the transparent blue lines. It is difficult to distinguish the individual error signals here as there was little difference between the scans. 
Here, the horizontal axis represents the frequency offset in Hz from the error point, measured in terms of the frequency difference between the fields, and the vertical axis is the error signal in V. The orange line shows a representation of the median slope obtained at the error point ($m_{\rm ES}$) of $(3.69\pm0.07)\times10^{-4}\rm\,V/Hz$. This was the value of $m_{\rm ES}$ used to calibrate the measurement of the out-of-loop noise for the \ac{PDH} stabilization loop of $E_{\parallel_+}$ with Equation~\ref{Eq:OOL}.

\end{document}